\documentclass[epj]{svjour}

\usepackage{graphicx}% Include figure files
\usepackage{bm}% bold math

\usepackage{epsfig}
\usepackage{graphics}
\usepackage{graphicx}
\usepackage{amssymb}

\usepackage{color}
\newenvironment{ckomm}{}{ }
\newcommand{\QQ}{\begin{ckomm}\color{red} \bf}
\newcommand{\QQEND}{\end{ckomm}}
\begin{document}
\title{A moving magnetic mirror to slow down a bunch of
atoms}
%generate a slow continuous beam}
\titlerunning{A moving magnetic mirror to ...}

\author{G. Reinaudi, Z. Wang, A. Couvert, T. Lahaye and D. Gu\'ery-Odelin}
\institute{Laboratoire Kastler Brossel\thanks{Unit\'e de Recherche
de l'Ecole Normale Sup\'erieure et de l'Universit\'e Pierre et
Marie Curie, associ\'ee au CNRS.
}, D\'epartement de Physique de l'Ecole Normale Sup\'{e}rieure, \\
24 rue Lhomond, 75005 Paris, France}

\date{\today}

\abstract{A fast packet of cold atoms is coupled into a magnetic
guide and subsequently slowed down by reflection on a magnetic
potential barrier ('mirror') moving along the guide. A detailed
characterization of the resulting decelerated packet is performed.
We show also how this technique can be used to generate a
continuous and intense flux of slow, magnetically guided atoms.}


\maketitle

\section{Introduction}
\label{sec:intro}

Ultracold and slow beams have a huge potential  in metrology,
matter wave interferometry and nanolithography
\cite{Interferometry}. The achievement of an ultracold beam of
neutrons  was made possible by specularly reflecting neutrons from
a Ni-surface which moved along the beam direction \cite{neutron}.
This breakthrough was followed by many achievements, among which
neutron optics experiments \cite{Werner} and neutron
interferometry experiments \cite{Rauch}.

In another context, the deceleration of a pulsed supersonic helium
beam through reflection on crystalline atomic mirrors mounted on a
high speed spinning rotor is currently being
investigated~\cite{raizen}, with also the goal of reaching a very
intense source for atom optics experiments.

 In this paper, we report on the slowing down
of packets of atoms injected into a magnetic guide through their
specular reflection from a moving magnetic mirror, and we also
demonstrate the use of this technique to generate a high flux of
slow and cold magnetically guided atoms. Such a beam is of crucial
importance in order to realize a cw atom laser by implementing the
forced evaporative cooling technique on a magnetically guided
beam, as proposed in Ref.~\cite{Mandonnet00} and experimentally
investigated in~\cite{lahaye05}.

 In the
reference frame of the mirror, the atomic velocity simply changes
sign after the reflection, going from $v_{\rm i}-v_{\rm m}$ to
$v_{\rm m}-v_{\rm i}$, where $v_{\rm i}$ is the relatively high
mean initial velocity of the packet and $v_{\rm m}$ is the mirror
velocity. In the laboratory frame, the final velocity of the
packet is thus:
\begin{equation}
v_{\rm f}=2v_{\rm m}-v_{\rm i}\,. \label{eq:finalv}
\end{equation}

Magnetic mirrors are based on the Zeeman interaction between an
inhomogeneous magnetic field and the atomic magnetic dipole
moment. Atoms in the low-field-seeking state are then reflected
elastically from high magnetic field regions. Magnetic mirrors for
atom optics have been implemented in various ways, using floppy
disks~\cite{floppydisk}, videotapes~\cite{videotape}, permanent
magnets arrays~\cite{sidorov} or
micro-electromagnets~\cite{microelectromagnet}.

Moving mirrors for cold atoms have been studied up to now using a
time-modulated, blue-detuned evanescent light wave propagating
along the surface of a glass prism. This technique has been used
to demonstrate atom optics manipulation: focussing of atomic
trajectories, formation of multiple images of a point source, and
phase-modulation of de Broglie waves~\cite{evanescentmirror}.

This paper is organized as follows. In section \ref{sec:setup} we
describe the experimental setup. In section \ref{sec:data} we
present our data on the slowing down of a single atomic packet
through its collision with the moving mirror. In section
\ref{sec:analysis} we detail the model that we have developed to
analyse our experimental data. Finally, we discuss our preliminary
results on the generation of a very slow and intense atomic beam
using this technique.

\section{Experimental setup}
\label{sec:setup}

\begin{figure}
\includegraphics[width=0.45\textwidth]{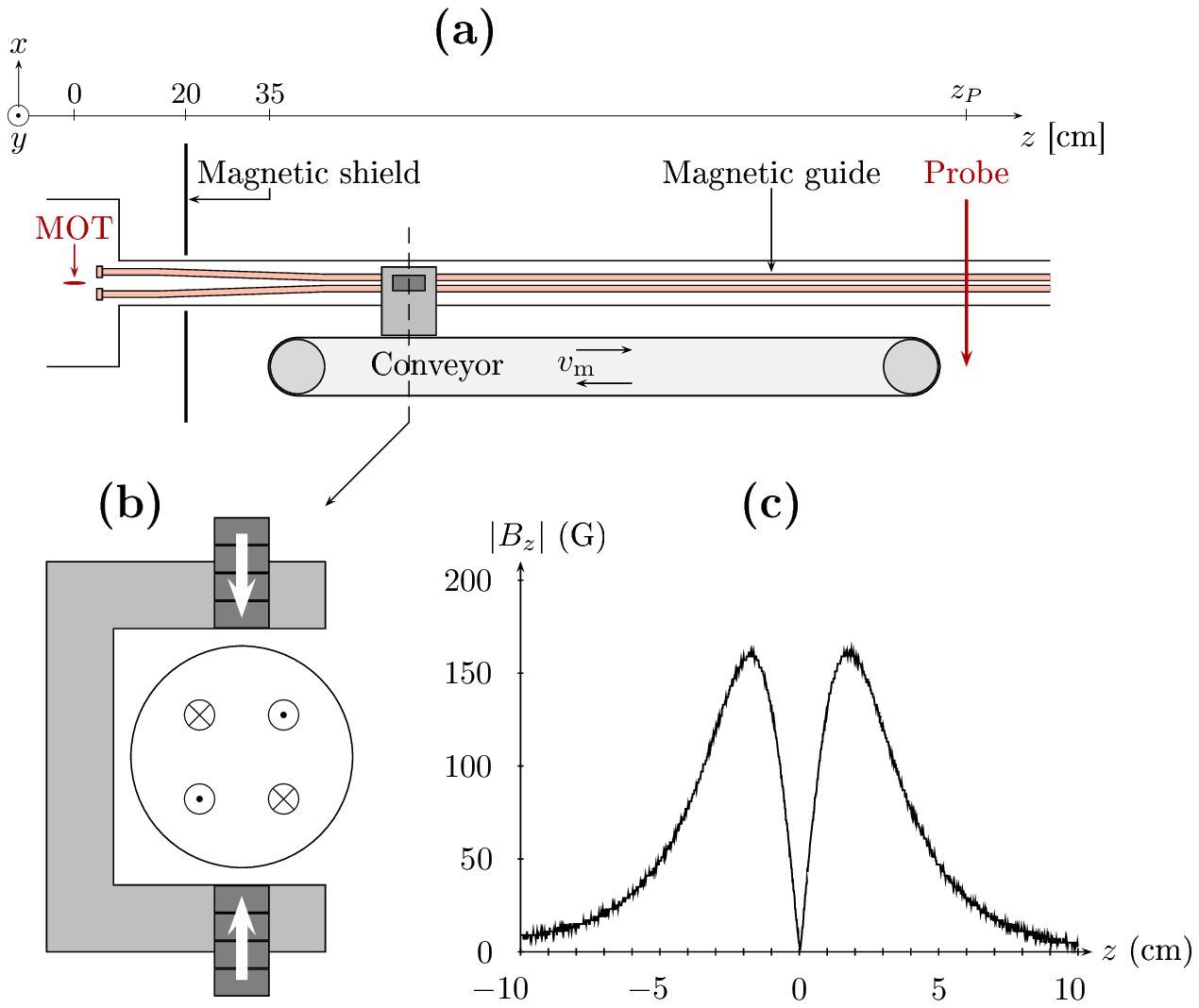}
\caption{(a) Overview of the experimental setup. (b) Transverse
cut that illustrates the 4 copper tubes of the magnetic guide in
the UHV chamber, in presence of the magnetic mirror generated by a
pair of magnets facing each other with opposite magnetization. (c)
Modulus of the longitudinal (along the $z$ axis) field generated
by the magnets.} \label{fig:setup}
\end{figure}

The experimental setup is schematically illustrated in Fig.
\ref{fig:setup}a (see also \cite{lahaye05}).

Atomic packets of a few $10^9$ cold $^{87}$Rb atoms are prepared
in a magneto-optical trap fed by a Zeeman slower. The cloud is
then set in motion at an adjustable velocity $v_{\rm i}$ by the
moving molasses technique, and optically pumped into the
weak-field-seeking ground state $|F=1,m_F=-1\rangle$ with magnetic
moment $\mu=\mu_B/2$ where $\mu_B$ is the Bohr magneton. It is
launched towards the entrance of a 4.5 meters long magnetic guide.
This guide, placed inside an ultra high vacuum chamber, consists
of four parallel water-cooled copper tubes (see Figs.
\ref{fig:setup}a and \ref{fig:setup}b). For the results given
below, a current $I=200$ A per tube was used, generating a
transverse magnetic field $B_{\rm g}=(bx,-by,0)$ with a gradient
of $b=500$ G/cm. At the beginning of the magnetic guide, the
packet is slowed down by undergoing a collision with a moving
magnetic mirror.

This mirror is made of a pair of rare-earth (Nd-Fe-B) permanent
magnets placed on a U-shaped support (see Fig. \ref{fig:setup}b).
The field magnitude controls the reflection of atoms in the
adiabatic regime when the magnetic moment follows the local field
direction. The chosen configuration with magnets facing each other
with an opposite magnetization provides a longitudinal potential
hill of height $\mu B_{\rm max}$, with $B_{\rm max}\simeq 160$ G.
This symmetric configuration minimizes the transverse magnetic
fields, thus avoiding a transverse deflection of the trajectories
during the reflection. We have plotted on Fig.~\ref{fig:setup}c
the measured absolute value of the longitudinal magnetic field
component of the mirror.

The transverse potential experienced by the atoms in the magnetic
guide is linear far away from the mirror as a result of the
two-dimensional quadrupolar configuration chosen to confine
transversally the atoms. The contribution of the mirror to the
longitudinal field tends mainly to smooth out the bottom of the
potential, resulting in a transverse harmonic confinement, while
essentially not modifying the transverse gradient provided by the
magnetic guide (the correction is less than 5\% for our range of
parameters). The magnetic mirror is fixed on a 1.2 meter long
conveyor belt parallel to the guide axis, which allows to control
its velocity $v_{\rm m}$ in a range of 20 cm/s to 120 cm/s. The
day-to-day velocity fluctuations are less than 2\%. In order to
control the position of the collision between the packet and the
moving magnetic mirror, we synchronize the launching of the cloud
into the guide with respect to the mirror motion.

The MOT region is protected by a magnetic shield from the
influence of the magnetic mirror. Atoms enter the conveyor area
after a propagation of $\sim 20$~cm into the guide. We measure the
characteristic of a given packet of atoms by monitoring the
absorption of a probe beam positioned at $z_P=1.75$ m (unless
otherwise stated) from the MOT location (see Fig. 1a). To detect
the atoms, we scan, 70 times per second, the frequency of the
probe across the $|5^2{\rm S}_{1/2},F=1\rangle\to|5^2{\rm
P}_{3/2},F'=0,1,2\rangle$ open transitions, at a rate of about
$600$~MHz/ms. We monitor the maximum of this absorption spectrum
with a sample-and-hold electronic circuit, which generates a data
point for the absorption every 15~ms. This technique reveals to be
remarkably insensitive to the inhomogeneity of the magnetic field
in the probe region, leading to a reliable and robust signal
proportional to the local atomic density~\cite{lahaye05}.

\section{Experimental results}
\label{sec:data}

\begin{figure}
\includegraphics[width=0.45\textwidth]{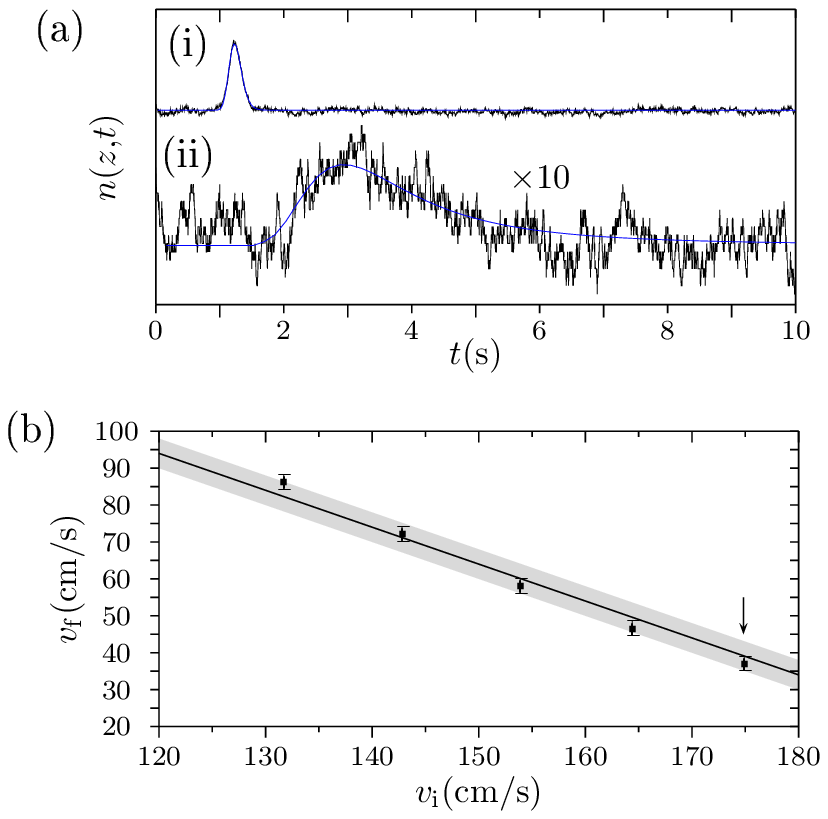}
\caption{(a) (i) Absorption signal from the probe placed at
1.75~meter from the MOT location. The time origin corresponds to
the launching of the packet. The injection velocity of the packet
is 142~cm/s. (ii) Absorption signal for a packet prepared in the
same conditions, but that has undergone a collision with the
magnetic mirror moving at a velocity $v_{\rm m } = 86\pm 2$ cm/s.
(b) Measured final velocity of the packet as a function of the
injection velocity for a mirror velocity of $107\pm 2 $ cm/s. The
straight line corresponds to the prediction of equation
(\ref{eq:finalv}), the grey area accounts for the mirror velocity
uncertainty. The arrow in (b) corresponds to parameters for which
95 \% of the initial kinetic energy of the packet has been removed
through the reflection on the mirror.} \label{data}
\end{figure}

When the relative velocity of the atoms with respect to the mirror
is larger than $\sim (2\mu B_{\rm max}/m)^{1/2}$, atoms pass over
the potential hill. We have indeed observed this effect by
launching a packet of atoms at this critical velocity on a
motionless mirror. Due to the dispersion of longitudinal
velocities in the packet, the packet was split into reflected and
transmitted ones, in good agreement with the measured height of
the magnetic potential barrier.

The absorption time-of-flight signal shown in Fig.~\ref{data}a~(i)
corresponds to a packet injected into the magnetic guide at a
velocity $v_{\rm i} = 142$ cm/s with an initial longitudinal
velocity dispersion of $\Delta v = 12$ cm/s with no mirror
present. The signal obtained after the collision with the mirror
moving at a velocity $v_{\rm m}=86 \pm 2$ cm/s is depicted on Fig.
\ref{data}a (ii). The packet of atoms has clearly been slowed
down. Indeed the measured final mean velocity is $\sim 35$~cm/s,
in good agreement with (\ref{eq:finalv}). For this specific
example, the maximum of the absorption peak is delayed by
approximately two seconds. The packet had consequently more time
to spread out which justifies the increase of its width and the
decrease of its height.

We have performed this experiment for different values of the
initial velocities $v_{\rm i}$ and for two values of the mirror
velocity ($v_{\rm m}=86\pm 2$ cm/s and $107\pm 2 $ cm/s). The
range of velocities that we have investigated was essentially
dictated by two constraints: (i) The finite barrier height $\mu
B_{\rm max}$, which limits the range of relative velocities
between the packet and the mirror that can be studied, (ii) The
length over which the collision occurs, which is limited by the
size of the conveyor. In addition, the finite lifetime of the
atoms due to collisions with the residual gas, combined with the
sensitivity of the detection, limits ultimately the range of
parameters over which reliable data can be obtained.

The measured final velocity $v_{\rm f}$ of the packet after its
collision with the moving mirror is shown on Fig. \ref{data}b as a
function of its injection velocity $v_{\rm i}$ for a mirror
velocity $107\pm 2 $ cm/s. We have been able to slow packets with
an initial velocity of 175~cm/s down to 35~cm/s. Through the
collision, we thus remove up to 95\% of the longitudinal kinetic
energy (experimental point indicated by a vertical arrow).

\section{Data analysis}
\label{sec:analysis}

The solid and smooth line superimposed on the experimental data
(see Fig.~\ref{data}a) results from a simple analytical model that
provides a good quantitative understanding of the collision
between the packet and the moving magnetic mirror, as well as a
flexible tool to analyze our experimental data.

In this model, we assume that the spatial distribution of the atoms
is initially a Dirac function. This assumption is reasonable in the
range of parameters experimentally investigated, since the size of
the packet during the collision is dictated essentially by the
initial velocity dispersion.

\begin{figure}
\includegraphics[width=0.45\textwidth]{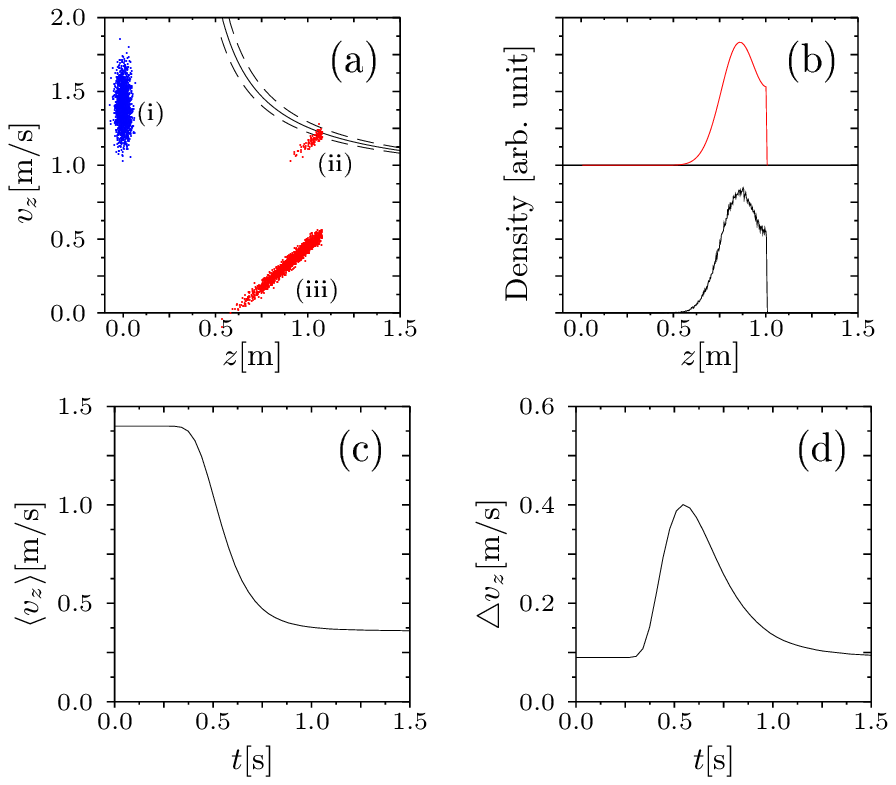}
\caption{Collision of a packet having an initial velocity $v_{\rm
i}=142$ cm/s and a velocity dispersion $\Delta v_z = 12$ cm/s with
an infinite repulsive wall moving at a velocity $v_m=88$ cm/s.
Initially the mirror is at a distance of $d=0.3$~m from the
initial mean position of the packet ($z=0$). (a) Atomic
distribution in the phase-space (i) at time $t=0$, and at time
$t=0.884$ s for (ii) and (iii). Atoms that belong to the class
(ii) have not yet been reflected, while the atoms from the class
(iii) have been slowed down through their collision with the
moving mirror. The curve in solid line represents the position
where the collision occurs for a given initial velocity assuming
that the initial position of the atom is $z=0$. The zone delimited
by the dashed lines accounts for the initial dispersion of the
size of the packet. (b) Atomic density profile of the packet after
a propagation time $t=0.884$ s. Upper graph: our model. Lower
graph: Monte Carlo simulation. (c) Mean velocity of the packet as
a function of time. (d) Velocity dispersion as a function of
time.} \label{num}
\end{figure}

The initial joint distribution in position and velocity of the
model is thus taken in the form: $\pi_0(z,v_z)=N\delta(z)p_0(v_z)$
where $p_0(v_z)$ is the velocity distribution. In practice we use
a gaussian function normalized to unity with a center velocity
$v_{\rm i}$, and a velocity dispersion $\Delta v$. The magnetic
mirror is modeled by an infinite repulsive potential wall. If the
mirror is initially positioned at $z=d$, a particle initially
positioned at $z_i<d$ with initial velocity $v_z$ is reflected
after a free propagation over a distance
$z^*=v_z(d-z_i)/(v_z-v_{\rm m})$, at time $t^*=(d-z_i)/(v_z-v_{\rm
m})$, and its velocity after the collision is, from
(\ref{eq:finalv}), $v'_z=2v_{\rm m}-v_z$. In Fig. \ref{num}a, we
provide an example of the packet evolution through a plot in the
phase-space $(z,v_z)$. We have plotted the results of a Monte
Carlo numerical simulation with $d=0.3$~m at times $t=0$ and at
$t=0.884$~s where 93\% of the atoms have already been reflected.
In this example, the packet (i) has an initial velocity of
$\langle v_z\rangle=142$ cm/s, a velocity dispersion $\Delta v_z =
12$ cm/s, and a dispersion in position $\Delta r=0.02$ m. Those
parameters are the ones of the experimental data of
Fig.~\ref{data}a. The density profile at time $t=0.884$~s given by
the numerical simulation performed with the real initial size, is
in excellent agreement with our simple analytical model
predictions without any adjustable parameter as shown on
Fig.~\ref{num}b.

Position-velocity correlations build up during the propagation:
particles with a larger initial velocity are reflected earlier,
which results in the inclined elliptical shape of the cloud in the
phase-space plot. The reflection on the mirror acts to some extent
as a translation in phase space while keeping the volume constant
as expected from Liouville's theorem \cite{Goldstein}. In Fig.
\ref{num}a atoms of class (ii) belong to the low velocity tail of
the initial distribution and have not yet been reflected, while
atoms of class (iii) have been significantly slowed down through
their collision with the moving mirror.

The density profile as a function of position and time is derived
from our model:
\begin{eqnarray}
n(z,t) & = & N\frac{\Theta(v_{\rm m}t+d-z)}{t}\bigg[
p_0\left(\frac{z}{t}\right)
\nonumber\\
&&+p_0\left(2v_{\rm m}+\frac{2d-z}{t}\right) \bigg].
\end{eqnarray}
The Heaviside step function $\Theta$ in the prefactor means that
all particles are always located before the mirror. The first term
is the contribution of non reflected atoms, while the second
accounts for the reflected particles.

This formula has been used to analyze our experimental data with a
gaussian form for $p_0$. The agreement is very good with all our
set of data. The value of the initial velocity and of the velocity
dispersion are deduced from the propagation of the packet with no
mirror present. We then plug the obtained values into our model to
deduce the characteristics of the slowed packet after its
interaction with the moving mirror (see Fig.~\ref{data}a). In
practice, we measure the instantaneous velocity of the conveyor
belt (i.e. of the mirror), and the adjustable parameters for the
fit are the final velocity, the amplitude of the absorption signal
and an effective initial position of the mirror which accounts for
the smoothness of the real mirror.

The evolution with time of the velocity distribution is obtained
by separating the contribution from atoms that have not yet
undergone a reflection, with the ones that have. At a given time
$t$ after the launch of the packet, the atoms with a velocity
larger than $\tilde{v}=v_{\rm m}+d/t$ have been reflected. We
deduce from our model the expression for the instantaneous
velocity distribution $p(v_z,t)$:
$p(v_z,t)=p_0(v_z)\Theta(\tilde{v}-v_z)+p_0(2v_{\rm
m}-v_z)\Theta(2v_{\rm m}-\tilde{v}-v_z)$. Starting with this
relation, one readily obtains the expression of the mean velocity
$\langle v_z \rangle(t)$, and of the dispersion $ \Delta v_z (t)$.
We plot on Figs.~\ref{num}c and \ref{num}d those quantities for
the experimental parameters of Fig.~\ref{data}a. The mean velocity
decrease occurs on a relatively short time scale. The velocity
dispersion turns out to not be a good quantity to characterize the
packet during the collision. Indeed, the coexistence of slowed and
fast particles results in a transient artificial increase of the
dispersion. However, an important feature is that the velocity
dispersion recovers its initial value after the collision with the
moving mirror.

This technique is intrinsically pulsed and this is the reason why
it has to be contrasted with other methods that can be used to
slow down a beam of atoms using an upward slope or/and a tapered
section with an increasing strength of the transverse confinement
over a portion of the guide. In the latter techniques, the
longitudinal velocity dispersion increases, while the mean
velocity decreases~\cite{cnsns} since the Liouville theorem
applied to a continuous beam dictates in one-dimension the
conservation of the product $\bar{v}\Delta v$ of the mean velocity
of the beam by its dispersion. The thermalization time between
transverse and longitudinal degrees of freedom is then drastically
increased by this large mismatch in dispersion velocities.

Another interesting outcome of the model is the distance over
which the collision occurs. To calculate the duration $\tau(\xi)$
of the collision, we determine the time $t^+(\xi)$ (resp.
$t^-(\xi)$) at which a given atom of initial velocity $v_{\rm
i}-\xi\Delta v$ (resp. $v_{\rm i}+\xi\Delta v$) collides with the
moving magnetic wall. We find:
\begin{equation}
\tau(\xi)=t^+(\xi)-t^-(\xi)=\frac{2\xi d \Delta v}{(v_{\rm
i}-v_{\rm m})^2-\xi^2\Delta v^2}\label{tau}
\end{equation}
The distance over which the collision occurs is then
$\delta(\xi)=v_{\rm m}\tau(\xi)$. For the parameters of Fig.
\ref{data}, $v_{\rm i}=142$ cm/s, $v_{\rm m}=88$ cm/s and $\Delta
v=12$ cm/s, we obtain $\tau\simeq 0.4$~s and $\delta \simeq
0.35$~m, with $\xi=1.45$ corresponding to 85\% percent of the
atoms reflected. This value is in good agreement with the duration
over which the mean velocity changes as illustrated in
Fig.~\ref{num}c.

\section{Generation of a continuous beam}
\label{sec:beam}

In this section we first summarize the methods that have been
demonstrated so far to produce a continuous and magnetically
guided atomic beam. We conclude with our preliminary results
towards this goal using a moving magnetic mirror, and the
improvements that this technique allows.

The realization of a magnetically guided beam has been first
achieved by {\it continuously} injecting atoms into a magnetic
guide using the moving molasses technique~\cite{cren,raithel}. An
improvement on the flux by nearly two orders of magnitude
($7\times 10^9$ atoms/s) with respect to this latter scheme was
obtained by feeding the magnetic guide \emph{periodically} at a
high repetition rate~\cite{prlrb2}. In this scheme, a cold packet
of atoms is loaded in an elongated magneto-optical trap (MOT) on a
short time $\tau_{\rm feed}\sim 100$~ms. It is then launched into
the magnetic guide and after a time delay $\tau_{\rm delay}\sim
100-200$~ms another packet is prepared. This delay ensures that
the light scattered from the MOT during the preparation of the new
packet does not generate too many losses on the previous one. The
guide is consequently fed at a rate of $1/(\tau_{\rm
feed}+\tau_{\rm delay})\sim 4\pm1$ packets per second. The
successive atomic packets then spread according to their
dispersion of longitudinal velocities, and eventually overlap. The
propagation time required to reach degeneracy through evaporative
cooling  is then dictated by the kinetics of evaporation in this
context~\cite{lahaye2006}.

Actually, the delay time $\tau_{\rm delay}$ is set by the
injection velocity. A longer propagation time in the guide
requires a priori an injection at lower velocity, which can be
done only at the expense of an increase of $\tau_{\rm delay}$, and
therefore of a reduction of the flux. To overcome this limitation,
two strategies have been proposed in~\cite{cnsns} and
experimentally demonstrated in~\cite{lahaye05}. They consist in
slowing down the beam, either by increasing the strength of the
transverse confinement with a tapered section, or by implementing
an upward slope in the guide.

By slowing down of packets of atoms injected into a magnetic guide
through their reflection on a moving magnetic mirror, one can use
a relatively high injection velocity $v_{\rm i}$, which permits in
turn to decrease $\tau_{\rm delay}$ when the sequence is repeated.
Consequently the flux coupled to the magnetic guide can be
increased.

\begin{figure}
\includegraphics[width=0.45\textwidth]{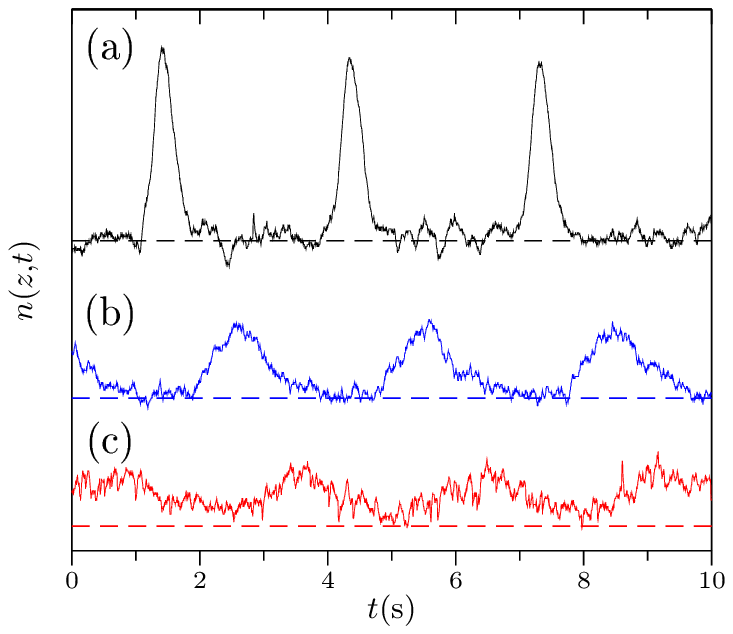}
\caption{Multiple injection and slowing of packets of atoms in the
magnetic guide. Packets with $v_{\rm i}=120$ cm/s probed after a
1.75 meter propagation in the magnetic guide : (a) in absence of
the moving mirror, (b) with a mirror set at a velocity $v_{\rm
m}=85$ cm/s which allows to reduce the velocity down to 50 cm/s.
(c) Same conditions as (b) but with a probe located at $z_P=2.25$
m. Curves have been shifted for clarity reasons, the dashed lines
correspond to a vanishing atomic density for each set of data.}
\label{fig:continuous}
\end{figure}

To demonstrate this new method to produce a continuous beam with a
very low mean velocity, we inject packets of atoms periodically
according to the motion of the mirror on the conveyor belt.
 We show in Fig. \ref{fig:continuous} the absorption
signal of the probe located in $z_P=1.75$ m for successive packets
launched into the magnetic guide at a velocity of $v_{\rm i}=120$
cm/s, in the absence of the moving mirror, and with a mirror set
at a velocity $v_{\rm m}=85$ cm/s. The final velocity of the
packet after it has undergone the collision is then 50~cm/s. With
a probe located 50~cm downstream the first probe we clearly see
the onset of overlapping between the successive slowed down
packets.

Equation (\ref{tau}) shows that the distance over which the
collision occurs can be considerably reduced by a proper choice of
the initial velocity of the packets and of the magnetic mirror.
For instance, with $v_{\rm i}=240$ cm/s and  $v_{\rm m}=140$ cm/s,
the collision occurs over a distance of the order of 10 cm. The
price to pay is in term of the barrier height that needs to be
larger than the kinetic energy of all the particles of the packet
for the mirror to be efficient for all atoms. With the latter
parameters, $B_{\rm max}
> m (v_{\rm i}-v_{\rm m}+4\Delta v)^2/(2\mu) \sim 340 $ Gauss. For
a given final velocity, the higher the injection velocity, the
more local the collision. In addition, the duty cycle in the
preparation of the packet can be significantly increased, and as a
direct consequence the flux also.

In conclusion, we have demonstrated the slowing down of large
packets of atoms propagating in a magnetic guide by reflection on
a moving magnetic mirror. We have also drawn the perspective of
this technique for the generation of a continuous, intense and
very slow beam of guided atoms.

\begin{acknowledgement}
We thank Jean Dalibard for a careful reading of the manuscript. We
acknowledge financial support from the D\'el\'egation G\'en\'erale
pour l'Armement (DGA). Our team belongs to the Institut Francilien
de Recherche sur les Atomes Froids (IFRAF). Z.~W. acknowledges
support from the European Marie Curie Grant MIF1-CT-2004-509423,
and G.~R. support from the DGA.
\end{acknowledgement}

\end{document}